\documentclass[aps,prd]{revtex4-1}
\usepackage{subfigure,graphics} 
\usepackage{amsmath, amsthm, amssymb}
\usepackage{flafter}

\begin{document} 

\title{
BRST-invariant boundary conditions and strong ellipticity} 
\author{Ian G. Moss}
\email{ian.moss@ncl.ac.uk}
\affiliation{School of Mathematics and Statistics, Newcastle University, 
Newcastle Upon Tyne, NE1 7RU, UK}

\date{\today}


\begin{abstract}
The quantisation of gauge theories usually procedes through the introduction of
ghost fields and BRST symmetry. In the case of quantum gravity
in the presence of boundaries, the BRST-invariant boundary value 
problem for the gauge field operators is non-elliptic, and consequently the definition of the 
effective action using heat-kernel techniques becomes problematic. 
This paper examines general classes of BRST-invariant boundary conditions and 
presents new boundary conditions for quantum gravity which fix the
extrinsic curvature on the boundary and lead to a well-defined effective action.
This prompts a discussion of the wider issue of non-ellipticity in BRST-invariant 
boundary value problems and when the use of gauge-fixing terms on the boundary
can resolve the issue.
\end{abstract}
\pacs{PACS number(s): }

\maketitle
\section{introduction}

A long-standing problem in quantum gravity has been the fact that the BRST-invariant boundary 
value problem is non-elliptic
\cite{Avramidi:1997sh,Esposito:1997wt,Avramidi:1997hy,Esposito:2004ts,Esposito:2005zg,Esposito:2012pi}.
The BRST-invariant boundary value problem is the very basic one in which the intrinsic metric is fixed and the DeWitt
gauge-fixing condition \cite{DeWitt67} is imposed at the boundary. Non-ellipticity implies that
these boundary conditions do not allow the construction of functional determinants by 
the usual techniques \cite{Barvinsky:1985an}. As a consequence, 
physical quantities, such as the scaling behaviour of coupling constants or coefficients
of anomalies, are potentially ill-defined. This paper explores the issue of non-ellipticity in 
BRST-invariant boundary value problems and shows that strong-ellipticity can be restored
in some cases by adding an extra boundary gauge-fixing term, an idea first
used by Barvinski \cite{Barvinsky:2006gd}. The boundary-value problem
for quantum gravity which fixes the extrinsic curvature at the boundary is an example.

Quantum effects are often described in terms of a one-loop effective action, (formally) given by
a series of terms of the form $\log\det P_m$, where $P_m$ are a set of second-order differential 
operators with an appropriate set of boundary conditions. The functional determinants can
be defined by heat kernels or zeta-functions of the operators, and the scaling behaviour of coupling 
constants is determined by terms in the the heat kernel asymptotics
\cite{Gilkey:1984,Moss:1989mz,Dowker:1995sw,Branson:1999jz,McAvity:1991xf,Vassilevich:2003xt}.

Boundary value problems have important applications in quantum field theory.
Possibly the earliest example was the derivation of the Casimir force between 
two parallel conducting plates in terms of quantum vacuum 
polarization \cite{Bordag:2001qi}.  In the 1980's, boundary value problems 
arose in quantum gravity in connection with the calculation of the Hartle-Hawking wave function of the universe 
\cite{Hartle:1983ai}, and later they appeared in the theory of strings and branes 
\cite{Tseytlin:1999dj,Barvinsky:2005ms,Barvinsky:2006gd}. 
Recently, boundary value problems have featured in the theory of brane cosmology \cite{Ahmed:2009ty} 
and membranes \cite{Berman:2009xd}. 

Early work on boundary value problems in quantum gravity 
\cite{Barvinsky:1987dg,Luckock:1989jr,Moss:1989wu,Luckock:1990xr}
indicated that the boundary conditions would be combinations of Dirichlet and Neumann (or Robin) acting
on different field components, which in the literature on quantum field theory are often called `mixed type'.
('Mixed type' is also used to refer to situations where Dirichlet and Neumann apply to different regions of the
boundary, but in this paper the same boundary conditions apply to the whole boundary).
In the general situation we have a field $\phi$ which is a section of a vector
bundle $V$ over a compact manifold with boundary. The vector bundle is decomposed 
at the boundary into $V=V_D\oplus V_N$ by  projection matrices $P_D$ and $P_N$.
Given a normal normal derivative $\nabla_n$ and an endomorphism $\psi$, 
then mixed boundary conditions are
\begin{equation}
P_D\phi=0,\quad (\nabla_n+\psi)P_N\phi=0.\label{mixedbc}
\end{equation}
If $\psi$ is replaced by a differential operator acting on the boundary, then we call these boundary 
conditions `mixed type with tangential derivatives'.

In BRST systems, there are ordinary fields $\phi$ and ghost fields $c$. Fields and ghosts 
are related by a BRST symmetry $s$, defined by a 
differential operator $D$,
\begin{equation}
 s\phi=D c.\label{sphi}
\end{equation}
We shall adopt the point of view that BRST invariance is an essential feature of quantum gauge theory
and that, in particular, the boundary conditions should be invariant under (\ref{sphi}) \cite{Moss:1996ip}
\footnote{Specifically, we require
the eigenvalue problem on the fields and the ghosts to be BRST-invariant}. 
We also restrict attention to mixed boundary conditions, with or without tangential derivatives. 
BSRT invariant boundary conditions of mixed type without tangential derivatives
have been found for Maxwell gauge theory and 
antisymmetric-tensor theory \cite{Moss:1990yq,Moss:1994jj}.

The BRST-invariant boundary conditions commonly used for quantum gravity are of mixed type with 
tangential derivatives \cite{Barvinsky:1987dg,Luckock:1990xr,Moss:1996ip,Esposito:1995zf}. 
These boundary conditions lead to a non-elliptic boundary
value problem when applied to the graviton operator
\cite{Avramidi:1997sh,Esposito:1997wt,Avramidi:1997hy,Esposito:2004ts,Esposito:2005zg,Esposito:2012pi}. 
The first step towards extracting physical predictions would
normally be to construct the heat kernel of the graviton operator, but this procedure is problematic when
faced with a non-elliptic system. This paper introduces a new set of BRST-invariant boundary 
conditions which fix the extrinsic curvature. These boundary conditions 
are strongly elliptic if we allow two tangential derivatives. The heat kernels, functional determinants
and effective action for these boundary conditions are all well-defined. (However,
Barvinsky et al. \cite{Barvinsky:2006pg} have pointed out that the small-time asymptotic expansion 
of the heat kernel depends crucially on the number of tangential derivatives). 

We shall see in the following sections that non-elliptic boundary value problems are typical of 
BRST-invariant boundary conditions. The main result of this paper is that, in many cases, the 
non-ellipticity can be explained as a residual gauge invariance in the system and it can be restored 
by allowing extra tangential derivatives in the boundary conditions. This has an interpretation in 
terms of adding extra gauge-fixing functions on the boundary. In certain limits, the heat kernel 
can be defined by omitting the residual gauge modes.

Calculations on spheres have suggested that the generalised zeta function of the operators 
relevant to quantum gravity could be constructed despite the non-ellipticity of the boundary value problem 
and divergences in the heat kernel \cite{Marachevsky:1995dr,Esposito:1995zf,Esposito:2012pi}.
It has also been suggested that the problems of non-ellipticity could be overcome by using 
non-Laplacian operators \cite{Avramidi:1997hy}, but it has not been shown that this can be done in a way
consistent with BRST symmetry. There are various techniques for dealing with specific non-elliptic 
boundary value problems
in other areas of mathematical physics 
\cite{MANA:MANA19800950130}.
One strategy involves imposing additional boundary conditions and relating the problem of interest
to an elliptic boundary value problem. The approach adopted here has some similarities, although
we work all the time under the restrictions of BRST symmetry and have to take this into account when
adding additional boundary conditions or gauge-fixing terms..

The plan of this paper is as follows. Section 2 gives a general account of BRST-invariant boundary conditions
for a one-dimensional system. Section 3 extends this to field theory, with explicit results for Maxwell
theory and linearised gravity. Section 4 gets to grip with the issue of non-ellipticity, its relation to residual
gauge freedoms and its resolution. General remarks are made in section 5.

The conventions used in this paper are as follows. Spacetime is replaced by a manifold with boundary,
on which there is a Riemannian metric and a Levi-Civita connection $\nabla_\mu$ or $;\mu$. On the 
boundary, the unit normal vector $n^\mu$ is ingoing and the induced connection $|i$.
Ordinary derivatives along the coordinate directions are denoted by $\partial_\mu$ or $\partial_i$.

\section{BRST-invariant boundary conditions}\label{general}

We begin with a general account of the BRST-invariant boundary conditions which will be used in 
subsequent sections. We shall follow the approach of Ref. \cite{Moss:1996ip}, including
details which were omitted in the earlier treatment.
The relevant features of BRST invariance can be described in a simple one-dimensional system 
with an abelian gauge symmetry. The boundary consists of just two points, representing the initial
time and the final time. The time coordinate will eventually become the normal 
coordinate to the boundary when we deal with a field theory which has additional spatial dimensions.

Consider the one dimensional system with coordinates $q=(q_i,q_a)$ and Lagrangian
\begin{equation}
L_q(q_i,\dot q_i,q_a).
\end{equation}
A Lagrangian of this form has a set of primary constraints $\pi_q^a=0$, where $\pi_q^a$ are the momenta conjugate to
$q_a$. The equations of motion $\dot\pi_q^a=0$ lead to a set of secondary constraints $E_q^a=0$, where
\begin{equation}
E_q^a=-{\partial L_q\over\partial q_a}.
\end{equation}
The gauge symmetry can be converted into a BRST symmetry $s$, with anti-commuting ghosts $c_a$. 
We take the BRST symmetry to be linear, with transformations
\begin{equation}
s q_i=\alpha_i{}^ac_a,\qquad s q_a=\dot c_a+\beta_a{}^b c_b,\label{brst}
\end{equation}
for some functions $\alpha_i{}^a$ and $\beta_a{}^b$ depending only on $t$. Gauge invariance of the action implies
that the Lagrangian transforms by a total derivative, which we can write as a term linear in $c_a$ 
with coefficient $\eta^a(q_i,q_a)$,
\begin{equation}
sL_q=(\eta^a c_a)\dot{}\,\,.
\end{equation}
If we just examine the $\dot c_a$ part of the BRST transformed Lagrangian $sL_q$ we find
\begin{equation}
E_q^a=\alpha_i{}^a\pi_q^i-\eta^a,\label{cons}
\end{equation}
where $\pi_q^i$ are the momenta conjugate to $q_i$. The BRST transformation of $\pi_q^i$
can be found by differentiating $sL_q$ with respect to $\dot q_i$,
\begin{equation}
s\pi_q^i={\partial\eta^a\over\partial q_i}c_a.\label{spi}
\end{equation}

The gauge symmetry is fixed by introducing auxiliary fields $b^a$ and gauge-fixing functions 
${f}_a(q,\dot q)$, which we take to be of the form
\begin{equation}
{f}_a=\dot q_a+\nu_a{}^i\dot q_i+\chi_a(q),\label{gff}
\end{equation}
where $\nu_a{}^i$ are functions of $t$.
The gauge-fixing and ghost Lagrangians are then
\begin{eqnarray}
L_{gf}&=&b^a\, {f}_a-\frac12\xi^{-1}G_{ab} b^ab^b,\\
L_{gh}&=&-\bar c^a\,s{f}_a+(\bar c^a \,sq_a)\dot{}\, ,\label{lgh}
\end{eqnarray}
where $\xi$ is a constant and $G_{ab}$ is a symmetric invertible matrix. 
The total derivative term is added to ensure that
the Lagrangian $L=L_q+L_{gf}+L_{gh}$ is first order in time-derivatives.
BRST symmetry is imposed by requiring that
\begin{equation}
s\bar c^a=b^a,\qquad s c_a=sb^a=0.\label{brstghost}
\end{equation}
Note that, due to the addition of the total derivite terms, the total Lagrangian $L$ transforms
under BRST by a total derivative, $sL=dj/dt$, where
\begin{equation}
j=\eta^ac_a+s\bar c^a\,sq_a.\label{defj}
\end{equation}

We are now ready to consider the BRST transformations of the full set of fields and their
conjugate momenta. The momenta obtained from the Lagrangian $L$ are
\begin{eqnarray}
\pi^i&=&\pi^i_q+\nu_a{}^ib^a,\label{mom1}\\
\pi^a&=&b^a,\label{pia}\\
\bar p_a&=&\dot c_a+\beta_a{}^b c_b,\label{pa}\\
p^a&=&-\dot{\bar c}+\bar c^b\left({\partial\chi_b\over \partial q_a}+\nu_b{}^i\alpha_i{}^a\right),
\label{mom4}
\end{eqnarray}
where the ghost momenta for $c_a$ and $\bar c^a$ are  denoted by $p^a$ and 
$\bar p_a$ respectively. The field transformations we have seen already,
\begin{eqnarray}
s q_i&=&\alpha_i{}^ac_a,\label{brst1}\\
s q_a&=&\bar p_a,\label{brst2}\\
s\bar c^a&=&b^a,\\
sc_a&=&sb^a=0.\label{brst4}
\end{eqnarray}
The momenta $\pi_q^i$ transform by Eq (\ref{spi}). The transformations of the remaining momenta
can be obtained directly from the definitions (\ref{mom1}-\ref{mom4}), 
\begin{eqnarray}
s\pi^i&=&{\partial\eta^a\over\partial q_i}c_a,\label{brstpi1}\\
sp^a&=&E^a+{\delta S\over \delta q_a},\label{brstpi2}\\
s\pi^a&=&s\bar p_a=0,\label{brstpi4}
\end{eqnarray}
where the constraint $E_q^a$ of Eq. (\ref{cons}) has been modified to include a $\nu_a{}^i$ term,
\begin{equation}
E^a=\alpha_i{}^a\pi^i-\eta^a.\label{defe}
\end{equation}
The transformations of the gauge-fixing function and the constraint are also useful,
\begin{eqnarray}
s{f}_a&=&-{\delta S\over \delta\bar c^a},\label{brstf}\\
sE^a&=&-{\partial\eta^a\over \partial q_b}\bar p_b.\label{brste}
\end{eqnarray}
The first of these expresssions is obtained by examination of the ghost Lagrangian (\ref{lgh})
and the second follows from Eqs. (\ref{defe}), (\ref{brstpi1}) and (\ref{brst2}).

We look for boundary conditions in configuration space which are BRST-invariant. 
The simplest possibility is a vanishing-ghost condition with $q_i$ fixed,
\begin{equation}
c_a=\bar c^a=b^a=0,\qquad q_i\hbox{ fixed}.
\end{equation}
These boundary conditions are standard in the BRST formalism, see for example \cite{Henneaux:1992ig}.
If we eliminate $b^a$ using the field equation $b^a=\xi G^{ab}{f}_b$ and use Eq. (\ref{gff}) for the 
gauge-fixing functions, then we arrive at a set of boundary conditions which we denote by ${\cal B}_M$,
\begin{equation}
{\cal B}_M:\quad c_a=\bar c^a=\dot q_a+\nu_a{}^i\dot q_i+\chi_a(q)=0,\qquad q_i\hbox{ fixed}.\label{mbc}
\end{equation}
The BRST invariance of the boundary conditions in this form is restricted due to the 
elimination of $b^a$. If we define the ghost operator $P_c$ by a functional derivative of the action $S$,
\begin{equation}
(P_c)_a{}^b={\delta^2 S\over\delta \bar c^a\,\delta c_b},
\end{equation}
then Eq. (\ref{brstf}) shows that the BRST variation of the gauge-fixing functional is
\begin{equation}
s{f}_a=-(P_c)_a{}^bc_b.
\end{equation}
Note that the boundary value problem for the eigenvalues of $P_c$ is still fully BRST-invariant, 
since $P_cc=\lambda c=0$ on the boundary. We shall describe a set of boundary conditions as BRST-invariant if
they are BRST-invariant for the eigenvalue problem. These boundary conditions can be used to
evaluate the functional determinants which appear in one loop quantum calculations.

The generalisation of the boundary conditions (\ref{mbc}) to electromagnetic field theory fixes the
magnetic field on a spacelike boundary. There are other possibilities, however, one being a vanishing-ghost 
condition with the momenta $\pi^i$ held fixed, which we denote by ${\cal B}_E$,
\begin{equation}
{\cal B}_E:\quad c_a=\bar c^a=\dot q_a+\nu_a{}^i\dot q_i+\chi_a(q)=0,\qquad \pi^i\hbox{ fixed}.\label{ebc}
\end{equation}
These correspond in electromagnetism to fixing the the electric field on a spacelike boundary. 
When we generalise to more than one dimension, these boundary conditions resemble the mixed type of 
boundary condition (\ref{mixedbc}) but in general the $\chi_a(q)$ terms introduce derivatives 
tangential to the boundary.

Another important set of boundary conditions can be obtained by having the ghost momenta vanish on the
boundary. In is then possible to require that the constraint function $E^a$ vanishes on the boundary since
its BRST variation vanishes by Eq (\ref{brste}). In a field theory context, the constraint function usually
has tangential derivatives. If we would like to remove as many tangential derivatives from the boundary 
conditions as possible, than it is useful to consider special cases where
\begin{equation}
\eta^a=\alpha_i{}^ah^i,\label{spece}
\end{equation}
for some function $f^i(q)$. This includes a fortiori the situation where the Lagrangian is BRST-invariant
and $\eta^i=0$. In these cases Eq. (\ref{defe}) reveals that $E^a=\alpha_i{}^a(\pi^i-h^i)$.
Fixing $E^a$ is then equivalent to the simpler condition of fixing $\pi^i-h^i$. Since $\alpha_i{}^a$
contain the tangential derivatives in the field theory case, this is an integration of the boundary conditions,
and Eq. (\ref{spece}) can be regarded as an integrability condition. The corresponding set of BRST-invariant 
boundary conditions is denoted by ${\cal B}'_E$,
\begin{equation}
{\cal B}'_E:\quad p^a=\bar p_a=q_a=0,\qquad \pi^i-h^i\hbox{ fixed},
\label{epbc}
\end{equation}
The BRST invariance is restricted because the field equation appears in the variation
of $p^a$ (see Eq. (\ref{brstpi2})),  but as before we have BRST  symmetry in the eigenvalue
problem. We will show later that for electromagnetism these boundary conditions are of 
mixed type with no tangential derivatives.

Many other sets of BRST-invariant boundary conditions can be constructed. The most general
set of linear boundary conditions can be found by starting from the basic set (\ref{mbc}) and
applying canonical transformations which preserve the form of the BRST operator \cite{Moss:1996ip},
\begin{equation}
\Omega=\bar p_a\pi^a+c_aE^a.\label{qdef}
\end{equation}
This leads to a family of boundary conditions which depends on matrices $B_a{}^b$, $D_a{}^i$ and $F_{ai}$,
\begin{eqnarray}
B_b{}^aE^a+D_b{}^a{f}^b&=&0,\label{bd}\\
B_b{}^ap^b+D_b{}^a\bar c^b&=&0,\\
B_a{}^b\bar p_b-D_a{}^b c_b&=&0,\\
B_a{}^bq_b+D_a{}^i q_i+F_{ai}\pi^i&&\hbox{ fixed}\label{bdf},
\end{eqnarray}
where $D_a{}^b=D_a{}^i\alpha_i{}^a-F_{ai}\partial\eta^b/\partial q_i$.
It  is simple to check directly that these form a BRST-invariant set, but when generalised to field theory
they usually have one or more tangential derivatives.

Up to now we have assumed that the Lagrangian $L_q$ is presented in a form which has primary
constraints $\pi^a_q=0$, but this is often not the case. Suppose, instead, that
a Lagrangian $L_q'$ has primary constraints $\pi^{a\prime}_q=h^a(q)$. In this situation,
we can introduce a canonical transformation generated by,
\begin{equation}
L'=L+\dot \epsilon,\label{can}
\end{equation}
where $\epsilon\equiv\epsilon(q_i,q_a.\bar c^a,c_a)$. The momenta for $L'$ and $L$ are
related by
\begin{eqnarray}
{\pi^i}'&=&\pi^i+{\partial\epsilon\over\partial q_i},\label{piprime}\\
{\pi^a}'&=&\pi^a+{\partial\epsilon\over\partial q_a},\\
{p^a}'&=&p^a+{\partial\epsilon\over\partial c_a},\\
{\bar p_a}'&=&\bar p_a+{\partial\epsilon\over\partial \bar c^a}.
\end{eqnarray}
Note that substituting these into the formula (\ref{qdef}) leaves the BRST operator unchanged. 
However, the boundary condition ${\pi^i}'=0$ is not BRST-invariant. 
What we will do in this case is determine the function $\epsilon$ which transforms the constraint into $\pi^a_q=0$,
\begin{equation}
{\partial\epsilon\over\partial q_a}={\partial L_q'\over\partial \dot q_a}.\label{defepsilon}
\end{equation}
The boundary condition (\ref{ebc}) becomes
\begin{eqnarray}
{\cal B}_E:\quad &&c_a=\bar c^a=\dot q_a+\nu_a{}^i\dot q_i+\chi_a(q)=0,\nonumber\\
&&{\pi^i}'-\partial\epsilon/\partial q_i\hbox{ fixed}.\label{enbc}
\end{eqnarray}
A similar canonical transformation can be used to add additional gauge-fixing functions
$F_a(q_i)$ on the boundary. This will prove useful later in the context of field theories. The new Lagrangian 
is defined as in Eq (\ref{can}) with extra terms
\begin{equation}
\epsilon=\frac12\alpha G_{ab}F_aF^b,
\end{equation}
Note that the field equations only depend on the value of $F_a$ at the boundary because
the Lagrangian is a function of $\dot\epsilon$.
The new momentum is obtained from Eq. (\ref{piprime}),
\begin{equation}
\pi^i=\pi^{i\prime}-\alpha G_{ab}{\partial F^a\over\partial q_i}F^b.
\end{equation}
The new set of BRST-invariant boundary conditions will be denoted by ${\cal B}_E(\alpha)$.
From Eq. (\ref{ebc}), these are
\begin{eqnarray}
{\cal B}_E(\alpha):\quad &&c_a=\bar c^a=\dot q_a+\nu_a{}^i\dot q_i+\chi_a(q)=0,\nonumber\\
&&{\pi^i}'-\alpha G_{ab} F^a\,\partial F^b/\partial q_i\hbox{ fixed}.\label{bbc}
\end{eqnarray}
The BRST symmetry can be shown directly using the transformations (\ref{brst1}-\ref{brstpi4}).

\section{Field theories}

In this section we shall consider the application of BRST boundary conditions to 
linearised gauge field theories. The gauge-invariant action for a set of fields $\varphi$ is of the form
\begin{equation}
S_\varphi=\int_{\cal M}d\mu\, {\cal L}_\varphi+\int_{\cal \partial M}d\mu\, {\cal L}_b,
\end{equation}
where ${\cal M}$ is a Riemannian manifold with metric $g_{\mu\nu}$, volume measure $d\mu$, 
Levy-Civita connection $\nabla_\mu$ and  boundary ${\cal\partial M}$. 
The quadratic action for the linearized fields arises from expanding of the fields about a 
background field configuration. This background can be chosen to satisfy an inhomogeneous 
boundary value problem for the classical field equations. We shall therefore focus on the homogenous
boundary value problems for the field fluctuations. The quadratic action defines a set of 
second order operators which depend on the background fields. The one-loop effective 
action is related to the functional determinants of these operators \cite{Schwinger,DeWitt1975295}.

We can set up a normal coordinate system close to the boundary with the coordinate $t$ 
along the unit normal direction and $t=0$ on the boundary. The eigenvalue problem for 
the fluctuation operators will be required to be BRST-invariant with boundary conditions 
of the mixed type described in the introduction.

\subsection{Maxwell theory}

The simplest example is provided by vacuum electrodynamics in curved space with Maxwell
field $A_\mu$ and field strength $F_{\mu\nu}=A_{\nu,\mu}-A_{\mu,\nu}$. The Lagrangian
density is the usual Maxwell form
\begin{equation}
{\cal L}_q=\frac14 F_{\mu\nu}F^{\mu\nu},
\end{equation}
with BRST symmetry
\begin{equation}
sA_\mu=c_{;\mu},
\end{equation}
and Lorentz gauge-fixing function
\begin{equation}
{f}=g^{\mu\nu}A_{\mu;\nu}.
\end{equation}
The total Lagrangian density is therefore
\begin{equation}
{\cal L}=\frac14 F_{\mu\nu}F^{\mu\nu}+bf-\frac12\xi^{-1}b^2+\bar c^{;\mu}c_{;\mu}.
\end{equation}
Under the BRST symmetry, $s{\cal L}_q=0$ with no boundary terms.

In order to set up the phase space near
the boundary ${\cal\partial M}$ we need to decompose the Maxwell field into
tangential and normal components $(A_i,A)$ (see appendix).
Decomposition of the Lagrangian density gives momenta
\begin{equation}
\pi^i=h^{ij}\left(\dot A_j-A_{|j}+K_j{}^k A_k\right),
\quad \pi=b.
\end{equation}
A similar reduction of the the gauge-fixing function gives
\begin{equation}
{f}=\dot A+KA+A_i{}^{|i}.
\end{equation}
The BRST generator is
\begin{equation}
\Omega=\bar p\,\pi+c\,\pi^i{}_{|i}
\end{equation}

The BRST boundary conditions are given in table \ref{table1}. The sets ${\cal B}_M$ and ${\cal B}'_E$ 
consist of mixed Dirichlet and Robin boundary conditions with no tangential derivatives, whilst the set ${\cal B}_E$
is first order in normal derivatives and contains tangential derivatives.
These BRST boundary conditions can be generalised to antisymmetric tensor fields \cite{Moss:1990yq,Moss:1994jj}, 
where the boundary  conditions ${\cal B}_M$ and ${\cal B}'_E$ are equivalent to the relative and 
absolute boundary conditions used in the index theory of the de Rahm complex \cite{Gilkey:1984}.

\begin{table}[htb]
\caption{\label{table1}Homogeneous BRST boundary conditions for Maxwell theory in curved space
with gauge-fixing function ${f}=A_\mu{}^{;\mu}$. The field $A_i$ is in the tangential direction, 
$A$ in the normal direction and the ghost is $c$.}
\begin{ruledtabular}
\renewcommand{\arraystretch}{2.5}
\begin{tabular}{llll}
Type&fixes&Dirichlet&non-Dirichlet\\
\hline
${\cal B}_M$&$F_{ij}$, ${f}$&$A_i=c=0$&$\dot A+KA=0$\\
${\cal B}_E$&$F_{in}$, ${f}$&$c=0$&$\dot A+KA+A_i{}^{|i}=0$\\
&&&$\dot A_i-A_{|i}+K_i{}^jA_j=0$\\
${\cal B}'_E$&$F_{in}$, $A$&A=0&$\dot A_i+K_i{}^jA_j=0$\\
&&&$\dot c=0$\\
\end{tabular}
\end{ruledtabular}
\end{table}

\subsection{Linearized gravity}

Linearized gravity forms the starting point for order $\hbar$ quantum gravity calculations
based on Einstein gravity, as well as having wider applications to supergravity, low energy
superstring theory and covariantly quantised strings and membranes. 
The Lagrangian density is obtained by decomposing the metric in the Einstein-Hilbert action
into a background $g_{\mu\nu}$ and a field $\gamma_{\mu\nu}$,
\begin{equation}
g_{\mu\nu}+2\kappa\gamma_{\mu\nu},
\end{equation}
where $\kappa^2=8\pi G$ and $G$ is Newton's constant in $m$ dimensions. 
We also use a dual field
\begin{equation}
{\overline\gamma}^{\mu\nu}=g^{(\mu\nu)(\rho\sigma)}\gamma_{\rho\sigma},
\end{equation}
defined by the DeWitt metric
\begin{equation}
g^{(\mu\nu)(\rho\sigma)}=\frac12\left(g^{\mu\rho}g^{\nu\sigma}+g^{\mu\sigma}g^{\nu\rho}
-g^{\mu\nu}g^{\rho\sigma}\right).
\end{equation}
Keeping only the quadratic terms and integrating by parts gives the Lagrangian 
\cite{DeWitt67,Barvinsky:1985an}
\begin{equation}
{\cal L'}_q=\frac12 {\overline\gamma}^{\mu\nu;\rho}\gamma_{\mu\nu;\rho}
-g^{\mu\nu}\overline\gamma_{\mu\rho}{}^{;\rho}\overline\gamma_{\nu\sigma}{}^{;\sigma}
+\frac12Q_{\mu\nu}{}^{\rho\sigma}\overline\gamma^{\mu\nu}\gamma_{\rho\sigma},
\end{equation}
where,
\begin{equation}
Q_{\mu\nu}{}^{\rho\sigma}=-2R_{(\mu}{}^{(\rho}{}_{\nu)}{}^{\sigma)}
-2\delta_{(\mu}{}^{(\rho}G{}_{\nu)}{}^{\sigma)}
+{2\over m-2}g_{\mu\nu}G^{\rho\sigma}+R_{\mu\nu}g^{\rho\sigma}.
\end{equation}
and $R_{\mu\nu\rho\sigma}$ is the Riemann tensor, $R_{\mu\nu}$ is the Ricci tensor 
and $G_{\mu\nu}$ is the Einstein tensor. 
The prime is used in the same way as in the previous section
to denote the fact that the primary constraint is not `$\pi^a=0$', and we will have to perform a canonical 
transformation to simplify the constraint before we can apply the boundary conditions ${\cal B}_E$.
 
The BRST symmetry transforms the metric fluctuations into a ghost field $c_\mu$,
\begin{equation}
s\gamma_{\mu\nu}= c_{\mu;\nu}+c_{\nu;\mu}.
\end{equation}
The Lagrangian density transforms into a total derivative term and, in the notation
of the previous section, the function $\eta$ is non-vanishing.
The commonly used gauge-fixing function is the DeWitt one,
\begin{equation}
{f}_\mu=\overline\gamma_{\mu\rho}{}^{;\rho}.
\end{equation}
The extra terms in the Lagrangian density from the gauge fixing are
\begin{equation}
{\cal L}_{gf+gh}=b^\mu f_\mu-\frac14\xi^{-1}b^\mu b_\mu
+\bar c^{\mu;\nu}c_{\mu;\nu}-R_\mu{}^\nu \bar c^\mu c_{\nu}.
\end{equation}
The Lagrangian has been put into in first-order form in order to apply the results of the previous section.
The graviton and ghost operators are obtained from the Lagrangian densities by
interation by parts after eliminating the auxilliary field $b^\mu$. The graviton operator is
\begin{equation}
P=-\delta_{\mu\nu}{}^{\rho\sigma}\nabla^2
+2(1-\xi)\delta_{\mu\nu}{}^{\alpha\delta}g^{(\beta\gamma)(\rho\sigma)}
g_{\gamma\delta}\nabla_\alpha\nabla_\beta
+Q_{\mu\nu}{}^{\rho\sigma},
\end{equation}
where $\delta_{\mu\nu}{}^{\rho\sigma}$ is the identity operator on symmetric tensors.
The gauge parameter $\xi$ has been scaled to make the graviton operator of Laplace type, 
i.e. with leading terms proportional to $\nabla^2$, when $\xi=1$

Near the boundary ${\cal\partial M}$ we need to decompose the metric fluctuation into
tangential and normal components $(\gamma_{ij},\gamma_i,\gamma)$ (see appendix).
We can identify $\gamma_{ij}$ with the physical variables $q_i$ of the previous section,
$\gamma_i$ and $\gamma$ with the constrained variables $q_a$. The momenta conjugate 
to $\gamma_i$ and $\gamma$ are
\begin{eqnarray}
{\pi_q^i}'&=&-2\gamma^{ij}{}_{|j}-2(K^{ij}-Kh^{ij})\gamma_j+\gamma^k{}_{k}{}^{|i}+\gamma^{|i},\\
{\pi_q}'&=&-\gamma_i{}^{|i}-K\gamma+K^{ij}\gamma_{ij},
\end{eqnarray}
where indices are raised using  the inverse boundary metric $h^{ij}$. These equations for the momenta become the 
primary constraints. The canonical transformation which trivialises these constraints corresponds to adding a term 
$\dot \epsilon$ to the Lagrangian density as in Eq (\ref{can}). Solving Eq. (\ref{defepsilon}) for $\epsilon$ gives
\begin{equation}
\epsilon=2\gamma^i\left(2\gamma_{ik}{}^{|k}-\gamma^k{}_{k|i}-\gamma_{|i}+
2K_i{}^k\gamma_k+2K\gamma_i\right)
+\frac12K\gamma^2-K^{ij}\gamma_{ij}\gamma.
\end{equation}
According to Eq. (\ref{piprime}), the momentum 
$\pi^{ij}=\partial{\cal L'}/\partial\dot\gamma_{ij}-\partial\epsilon/\partial \gamma_{ij}$  which appears in the
boundary condition ${\cal B}_E$ (see eq. (\ref{enbc})) is then
\begin{equation}
\pi^{ij}=\dot\gamma^{ij}-h^{ij}\dot\gamma^k{}_k-2\gamma^{(i|j)}-(K^{ij}-Kh^{ij})\gamma
+2h^{ij}\gamma_k{}^{|k}.\label{newpi}
\end{equation}
(Note that the convention here is to raise indices after taking the normal derivatives).

Unsurprisingly, the momentum $\pi^{ij}$ has a physical interpretation in terms of the canonical
decomposition of Einstein gravity. The Hamiltonian formalism gives the canonical momentum
\begin{equation}
p^{ij}={1\over 2\kappa^2}\left(K^{ij}-h^{ij}K\right).
\end{equation}
Let $\delta p^{ij}$ be the perturbation in the canonical momentum corresponding to
the metric perturbation $2\kappa\gamma_{\mu\nu}$,
then to first order one finds that
\begin{equation}
\delta p^{ij}={1\over 2\kappa}\pi^{ij},
\end{equation}
where  $\pi^{ij} $ is given by (\ref{newpi}). Boundary conditions on $\pi^{ij}$
correspond to fixing the canonical momentum on the boundary. This is analagous in the
Maxwell case to fixing the electric field on the boundary.

There are two basic sets of BRST-invariant boundary conditions, ${\cal B}_M$ corresponding to fixing the intrinsic
on the boundary \cite{Barvinsky:1987dg}, and the new boundary conditions ${\cal B}_E$ which fix
the extrinsic curvature. Boundary conditions of type ${\cal B}'_E$ only exist when the integrability 
condition Eq. (\ref{spece}) is satisfied,
We shall see later that this restricts the extrinsic curvature to the special form 
$K_{ij}=\kappa h_{ij}$, for some $\kappa$.
Only the boundary conditons for linearised gravity which apply for a general background are listed 
in table \ref{table2}.

\begin{table}[htb]
\caption{\label{table2}Homogeneous BRST boundary conditions for linearized gravity. 
The metric variation $\gamma_{\mu\nu}$ and ghost $c_\mu$
have been decomposed into tangential and normal components.}
\begin{ruledtabular}
\renewcommand{\arraystretch}{2.5}
\begin{tabular}{llll}
Type&fixes&Dirichlet&non-Dirichlet\\
\hline
${\cal B}_M$&$h_{ij}$, ${f}_\mu$&$\gamma_{ij}=c=c_i=0$&$\dot\gamma_i-\frac12\gamma_{|i}+K_i{}^j\gamma_j+K\gamma_i
=0$\\
&&&$\dot\gamma-\dot\gamma_i{}^i+2\gamma_i{}^{|i}+2K\gamma=0$\\
${\cal B}_E$&$K_{ij}$, ${f}_\mu$&
$c=c_i=0$&$\dot\gamma_i-\frac12\gamma_{|i}+\gamma_{ij}{}^{|j}-\frac12\gamma^j{}_{j|i}
+K_i{}^j\gamma_j+K\gamma_i=0$\\
&&&$\dot\gamma+2K\gamma-2K^{ij}\gamma_{ij}=0$\\
&&&$\dot\gamma_{ij}-2\gamma_{(i|j)}-K_{ij}\gamma=0$\\
\end{tabular}
\end{ruledtabular}
\end{table}

\section{Restoring strong ellipticity to the boundary value problem}

We now turn to the heat kernel of the BRST boundary value problem and the important issue of strong ellipticity.
Avrimidi et al. \cite{Avramidi:1997sh} have shown that the gravitational boundary value problem with fixed boundary
metric and DeWitt gauge condition is not elliptic, and therefore the heat kernel and the propagator
are ill-defined. We would like to examine whether this is also the case for the new BRST
invariant gravitational boundary conditions.

Consider the eigenvalue problem
\begin{eqnarray}
P f&=&\lambda f\hbox{ on }{\cal M},\\
B f&=&0\hbox{ on }{\cal\partial M},\label{evp}
\end{eqnarray}
where $P$ is a second order operator and $B$ is first-order in normal derivatives.
Ellipticity can be defined in terms of the leading symbols of the operators. The symbol of the 
operator $P$, denoted by $\sigma(P, x,\zeta)$, is constructed by
replacing derivatives $\partial/\partial x^\mu$ by $i\zeta_\mu$. The leading symbol 
$\sigma_L(P, x,\zeta)$ is obtained by keeping only the leading terms in $\zeta$.
The operator is elliptic if $\det\sigma_L(P, x,\zeta)\ne 0$ for $\zeta\ne 0$.

For the boundary value problem, we replace the tangential components of $\zeta$ by $k_i$
and the normal component by $-i\partial_t$
and consider the leading order system,
\footnote{Strictly speaking, we introduce a grading on the bundle of fields according
to the number of normal derivatives, and keep the leading terms in each sector of
this grading.}
\begin{eqnarray}
\sigma_L(P,x,k_i,-i\partial_t)f_L&=&\lambda f_L,\quad t>0,\\
\sigma_L(B,x,k_i,-i\partial_t) f_L&=&0,\qquad t=0.\label{se}
\end{eqnarray}
The boundary value problem is said to be strongly elliptic in $C-R_+$ if the operator $P$ is elliptic and
there are no bounded exponential solutions $f_L$ with $\lambda\in C-R_+$. 
The boundary value problem is elliptic if there are no bounded exponential
solutions with $\lambda=0$. Strong ellipticity implies the existence of a complete set of 
eigenfunctions to the original boundary value problem. Their eigenvalues are real and can be placed in an 
unbounded sequence $\lambda_1\le\lambda_2\dots$.

The heat kernel $K(x,x',\tau,P,B)$  for $x,x'\in {\cal M}$ and $\tau>0$ is defined in 
terms of the normalised eigenfunctions $f_n$ by
\begin{equation}
K(x,x',\tau,P,B)=\sum_{n=1}^{\infty}f_n(x)f_n(x')^\dagger e^{-\lambda_n \tau}.
\end{equation}
The sum converges if the boundary value problem is strongly elliptic, but it might not
exist if the operator is not strongly elliptic. In the strongly-elliptic case, the integrated 
heat kernel can be written in the form,
\begin{equation}
{\rm Tr}_{\cal B}\left(e^{-P \tau}\right)=\int_{\cal M}K(x,x,\tau,P,B)\,d\mu,
\end{equation}
where $d\mu$ is the volume measure on the manifold. We also define the generalised
zeta function by
\begin{equation}
\zeta(s)={\sum_{n=1}^{\infty}}'\lambda_n^{-s},
\end{equation}
where the prime denotes the omission of any vanishing eigenvalues.

For the BRST boundary conditions we have considered so far, and with gauge parameter $\xi=1$, 
the equations for ellipticity become
\begin{eqnarray}
\left(-\partial_t^2+k^2\right)f_L&=&\lambda f_L,\quad t>0,\\
\left(P_D+P_N\partial_t+\Gamma(k_i)\right)f_L&=&0,\qquad t=0,\label{defgamma}
\end{eqnarray}
where $P_D$ and $P_N$ are the projection matrices defined in the introduction,
and $\Gamma(k_i)$ is a antihermitian matrix acting on $V_N$. 
The boundary value problem is strongly elliptic if the spectrum 
$\hbox{spec}(\Gamma)\subset (-\infty,k)$ \cite{Gilkey:1983xz}. (It can be shown
that the same condition applies for any gauge parameter $\xi>0$.)

The matrices $\Gamma(k_i)$ can be obtained from the boundary conditions in tables
\ref{table1} and \ref{table2}. We keep only the derivative terms and replace the
tangential derivatives with $ik_i$. It is convenient to use the spacetime components, for example
the Maxwell ${\cal B}_E$ boundary condition becomes
\begin{equation}
\sigma_L(B,x,k_i,-i\partial_t)A=
\dot A_\mu+in_\mu k^\nu A_\nu-ik_\mu n^\nu A_{\nu}
\end{equation}
where $A_{\mu}=A_i e^i{}_\mu+An_\mu$ and $k_{\mu}=k_i e^i{}_\mu$. (The tangent basis
$e_i{}^\mu$ is defined in the appendix). We can see from table \ref{table3} that in the cases with
tangential derivatives, $\Gamma(k)$ always has an eigenvector with eigenvalue $k$.
Therefore, of the five sets of BRST boundary conditions considered so far, only those 
without tangential derivatives are strongly elliptic. 

\begin{table}[htb]
\caption{\label{table3}The spectrum of the tangential term $\Gamma(k)$ in the boundary 
operator for different BRST boundary conditions. The eigenvector
in the table is the eigenvector with eigenvalue $k$ which makes the operator non-elliptic.}
\begin{ruledtabular}
\renewcommand{\arraystretch}{2.5}
\begin{tabular}{llllll}
Field&Type&$P_N$&$\Gamma(k)$&spectrum&$k$-eigenvector\\
\hline
Maxwell&${\cal B}_E$&$\delta_\mu{}^\nu$&$i(n_\mu k^\nu-k_\mu n^\nu)$&$0,\pm k$&$kn_\mu-ik_\mu$\\
Gravity&${\cal B}_M$&$2n_{(\mu} n^{(\rho}\delta_{\nu)}{}^{\sigma)}-n_\mu n_\nu n^\rho n^\sigma$&
$i(n_\mu k_\nu n^\rho n^\sigma-2n_\mu n_\nu k^\rho n^\sigma)$&
$0,\pm k$&$kn_\mu n_\nu+ik_\mu n_\nu$\\
Gravity&${\cal B}_E$&$\delta_{\mu\nu}{}^{\rho\sigma}$&
$i(2k_\mu h_\nu{}^\rho n^\sigma-2n_\mu h_\nu{}^\rho k^\sigma
-n_\mu k_\nu\delta^{\rho\sigma})$&
$0,\pm k$&$kn_\mu k_\nu-ik_\mu k_\nu$\\
\end{tabular}
\end{ruledtabular}
\end{table}

The explanation for this lack of strong
ellipticity can be seen when we look at bounded exponential solutions to the leading
order equations (\ref{se}) when $\lambda=0$,
\begin{equation}
f_L=u_L(k)e^{-kt},
\end{equation}
where $u_L$ is the eigenvector with eigenvalue $k$ given in the table. In each case,
at leading order, these are gauge transformations of the form $f=Dv$, i.e.
\begin{equation}
f_L=\sigma_L(D,x,k,-i\partial_t)v_L(k)e^{-kt},
\end{equation}
for some $v_L$. We should therefore examine whether these correspond to an
infinite set of zero modes $P f=0$. If so, there is a residual gauge invariance
which has not been fixed by the gauge-fixing condition. In the next section,
we shall see how adding a boundary term can fix the remaining gauge invariance
and restore strong ellipticity. The idea of using boundary gauge-fixing terms is due to
Barvinsky \cite{Barvinsky:2006gd}.

The extra gauge-fixing terms come with their own gauge parameter.
In certain limits of the gauge parameter, the gauge choice is frozen
and the effect on the heat kernel is equivalent to leaving out the
the gauge zero-modes. The heat kernel defined with only the non-zero
modes will be denoted in the usual way by a prime,
\begin{equation}
K'(x,x',\tau,P,B)={\sum_{n=1}^{\infty}}' f_n(x)f_n(x')^\dagger e^{-\lambda_n \tau}.
\end{equation}
This can be used to define the trace ${\rm Tr}'$ and the generalised zeta function by
\begin{equation}
\zeta(s)={1\over\Gamma(s)}\int_0^\infty d\tau \,\tau^{s-1}
{\rm Tr}^\prime_{\cal B}\left(e^{-P \tau}\right).
\end{equation}
The bounded exponential solutions might not correspond to gauge zero modes in
every case, and then the additional gauge-fixing will not help with the ellipticity
problems.

\subsection{Maxwell theory}

The boundary conditions ${\cal B}_E$ for Maxwell theory are the ones which lead
to a non-elliptic boundary value problem. We can try adding the boundary gauge-fixing
function $F(A_i)$ given by
\begin{equation}
F(A_i)=A_i{}^{|i}.
\end{equation}
The boundary conditions in table \ref{table2} are modified into the new
set ${\cal B}_E(\alpha)$ given by Eq. (\ref{bbc}),
\begin{eqnarray}
\dot A_i-A_{|i}+\alpha A_j{}^{|j}{}_i+K_i{}^jA_i&=&0,\label{max1}\\
\dot A+A_i{}^{|i}+KA&=&0.\label{max2}
\end{eqnarray}
Ellipticity depends on the matrix $\Gamma$, which can be read off (\ref{max1}) by 
replacing the tangential derivatives with $ik_i$ and comparing with Eq. (\ref{defgamma}).
The eigenvectors of $\Gamma$ with non-zero eigenvalue are given by two
free parameters $x$ and $y$ and have the form $A_i=x\hat k_i$ and $A=y$, so that the
eigenvalue problem for $\Gamma$ reduces to
\begin{equation}
\begin{pmatrix}
-\alpha k^2&ik\\
-ik&0\\
\end{pmatrix}
\begin{pmatrix}x\\y\\\end{pmatrix}=
\lambda \begin{pmatrix}x\\y\\\end{pmatrix}
\end{equation}
The condition for strong ellipticity is that the largest eigenvalue is less than $k$,
and this is the case for $\alpha>0$. (This is also the correct sign choice for
convergence of the functional integral in the quantum system). We could therefore 
use the boundary conditions (\ref{max1}) and (\ref{max2}) to provide a 
well-defined heat kernel which can be used in one-loop quantum calculations.
However, we shall see how it is possible to relate the boundary conditions 
${\cal B}_E(\alpha)$ to a simpler set ${\cal B}'_E$.

The $\alpha\to\infty$ limit projects out (at least formally) the transverse modes
with $F(A_i)=0$. This can be examined further by decomposing the modes using 
the exterior derivative $d$ and its  conjugate $d^\dagger$. The Maxwell field 
operator is
\begin{equation}
P=d^\dagger d+\xi d\,d^\dagger,
\end{equation}
Consider the pure gauge field
\begin{equation}
A=d\chi,\quad P_c\chi=0,\label{gauge}
\end{equation}
where $P_c=d^\dagger d$. Both $dA$ and $d^\dagger A$ vanish, and so this
field satisfies the boundary conditions ${\cal B}_E$.
The Maxwell field (\ref{gauge}) is also a zero mode of $P$. There are no restrictions on
$\chi$ at the boundary and an infinite set of gauge zero modes like this can be constructed.
Imposing the gauge condition $F(A_i)=0$ removes these modes.

In the previous section we defined the heat kernel $K'(x,x',t,P,B_E)$ by taking a sum
over the non-zero modes of $P$. Hodge decomposition gives two types of
non-zero modes which satisfy the boundary conditions ${\cal B}_E$,
\begin{enumerate}
\item Exact: $f_n=\lambda_n^{-1/2}d\chi_n$, where $P_c\,\chi_n=\lambda_n \chi_n$ and 
$\chi_n=0$ on ${\cal \partial M}$.
\item Co-closed: $d^\dagger f_n=0$, where $F_{in}=0$ on ${\cal\partial M}$.
\end{enumerate}
In the co-closed case, part of the boundary condition is redundant because it is enforced by the
Hodge decomposition. It is possible to relate this boundary value problem to the 
boundary value problem ${\cal B}'_E$ (see table \ref{table1}).
Again, we use a Hodge decomposition of the modes, but with boundary conditions ${\cal B}'_E$
\begin{enumerate}
\item Exact: $f_n=\lambda_n^{-1/2}d\chi_n$, where $P_c\,\chi_n=\lambda_n \chi_n$ and 
$\dot \chi_n=0$ on ${\cal \partial M}$.
\item Co-closed: $d^\dagger f_n=0$, where $F_{in}=0$ on ${\cal\partial M}$.
\end{enumerate}
The co-closed modes for the two sets of boundary conditions are identical, whereas the exact modes come
from Dirichlet scalars in the first case and Neumann scalars in the second. 
If we split the mode sums into exact and co-exact mode sums,
then $K'(x,x',t,P,B_E)$ is determined by the heat kernels of the 
strongly elliptic boundary value problems ${\cal B}'_E$, 
the scalar Dirichlet problem ${\cal B}_D$ and the scalar Neumann problem ${\cal B}_N$. 
For example,  the integrated kernel,
\begin{equation}
{\rm Tr}^\prime_{{\cal B}_E}\left(e^{-P \tau}\right)=
{\rm Tr}_{{\cal B}'_E}\left(e^{-P \tau}\right)-{\rm Tr}_{{\cal B}_N}\left(e^{-P_c \tau}\right)
+{\rm Tr}_{{\cal B}_D}\left(e^{-P_c \tau}\right).\label{trp}
\end{equation}
There is a nice physical interpretation of this result which arises when we combine the fields and the ghosts
together into the supertrace. Let $\Delta=(P,P_c)$, then
\begin{equation}
{\rm STr}_{\cal B}\left(e^{-\Delta \tau}\right)={\rm Tr}_{\cal B}\left(e^{-P_c \tau}\right)
-{\rm Tr}_{\cal B}\left(e^{-P_c \tau}\right)
\end{equation}
The supertrace determines the effective action of the quantum theory of gauge fields in the-one loop
approximation. The ghost boundary conditions in ${\cal B}_E$ are Dirichlet and the ghost 
boundary conditions in ${\cal B}'_E$ are Neumann. Using Eq. (\ref{trp}), we have
\begin{equation}
{\rm STr}'_{{\cal B}_E}\left(e^{-\Delta \tau}\right)={\rm STr}_{{\cal B}'_E}\left(e^{-\Delta \tau}\right).
\end{equation}
Therefore the one-loop effective action for the quantum field theory derived from the
non-elliptic BRST boundary conditions ${\cal B}_E$ by omitting the zero modes is the same as the 
one-loop effective action derived from the strongly-elliptic BRST boundary conditions ${\cal B'}_E$.

\subsection{Linearized gravity}

We shall repeat the preceding analysis for linearized gravity with the new BRST 
boundary conditions ${\cal B}_E$ which fix the extrinsic curvature at the boundary.
The boundary condition $\pi^{ij}=0$ can be replaced by
\begin{equation}
\pi^{ij}-2\alpha \left(F^{(i|j)}-h^{ij}F^k{}_{|k}\right)=0,
\end{equation}
where
\begin{equation}
F_i=\gamma_{ij}{}^{|j}-\frac12\gamma^j{}_{j|i}.
\end{equation}
In terms of the metric components, using Eqs. (\ref{newpi}), (\ref{gfi}) and (\ref{gf}),
\begin{eqnarray}
\dot\gamma_{ij}-2\gamma_{(i|j)}-K_{ij}\gamma-2\alpha F^{(i|j)}&=&0,\label{newg1}\\
\dot\gamma_i-\frac12\gamma_{|i}+\gamma_{ij}{}^{|j}-\frac12\gamma^j{}_{j|i}
+K_i{}^j\gamma_j+K\gamma_i&=&0,\label{newg2}\\
\dot\gamma+K\gamma-2K^{ij}\gamma_{ij}&=&0\label{newg3}.
\end{eqnarray}
We replace the tangential derivatives with $ik_i$ and collect them together
into the matrix $\Gamma$ as in Eq.  (\ref{defgamma}). The non-zero eigenvalues
of $\Gamma$ arise from eigenvectors of the form $\gamma_{ij}=x\hat k_i\hat k_j$
and $\gamma_i=y\hat k_i$. We have $F_i=k_ix/2$ and the eigenvalue
problem becomes
\begin{equation}
\begin{pmatrix}
-\alpha k^2&ik\\
-ik&0\\
\end{pmatrix}
\begin{pmatrix}x\\y\\\end{pmatrix}=
\lambda \begin{pmatrix}x\\y\\\end{pmatrix}
\end{equation}
This is identical to the eigenvalue problem for the Maxwell theory, and again
we have $\hbox{spec}(\Gamma)\subset (-\infty,k)$ and we conclude that
Eqs. (\ref{newg1}-\ref{newg3}) provide a well-posed boundary value problem
for the graviton operator when $\alpha>0$.

To analyse the residual symmetry when $\alpha=0$ it is convenient to use the
operators $D$ and $D^\dagger$ defined by
\begin{eqnarray}
Dc_{\mu\nu}&=&c_{\mu;\nu}+c_{\nu;\mu},\\
D^\dagger\gamma_\mu&=&\bar\gamma_{\mu\nu}{}^{;\nu}.
\end{eqnarray}
A pure gauge field $\gamma_{\mu\nu}=Dc_{\mu\nu}$ 
will satisfy the boundary conditions ${\cal B}_E$ if $s\pi_{ij}=0$ and $D^\dagger\gamma=0$.
The BRST variation of $\pi^{ij}$ using Eqs. (\ref{sg1}-\ref{sg3}) is
\begin{equation}
s\pi_{ij}=-c_{|ij}+h_{ij}c_{|k}{}^k-K^{(i}{}_kc^{j)|k}+K^{k(i}c_k{}^{|j)}.
\end{equation}
A sufficient condition for $s\pi_{ij}=0$ is that $c=0$ and $c_{[i|j]}=0$ on the boundary, 
leaving us free to pick one arbitrary function $a$ on the boundary,
where $c_i=a_{|i}$. The gauge mode will be a zero mode of gauge-fixed gravity 
operator $P$ if it satisfies the
gauge condition $D^\dagger\gamma=0$, i.e.
\begin{equation}
D^\dagger D c=0
\end{equation}
This is an inhomogeneous Dirichlet problem and we have a solution for each choice of the
arbitrary function $a$ on the boundary. These modes are responsible for the non-ellipticity.
They are eliminated by restricting the modes using the boundary gauge condition $F_a=0$, 
or equivalently by using the boundary value system Eqs. (\ref{newg1}-\ref{newg3}).

In the Maxwell case it was possible to relate the boundary conditions ${\cal B}_E$ to a strongly
elliptic BRST-invariant boundary value problem ${\cal B}'_E$ with no tangential derivatives. We can 
do this also for linearised gravity for a restricted class of backgrounds where the strongly 
elliptic BRST-invariant boundary value problem of this type exists. Consider, for example,
the case where the extrinsic curvature is proportional to the surface metric,
\begin{equation}
K_{ij}=\kappa h_{ij}
\end{equation}
The BRST-invariant boundary value problem for these backgrounds was found in Ref. \cite{Moss:1996ip},
\begin{equation}
{\cal B}'_E:\quad\dot\gamma_{ij}-K_{ij}\gamma=0,\,
\dot \gamma+K\gamma-K\gamma_i{}^i=0,\,
\gamma_i=0,\,
\dot c_i-K_i{}^jc_j=0,\,c=0.
\end{equation}
As in the Maxwell case, the two sets of boundary conditions can be related by dividing the modes into
the image of $D$ and the kernel of $D^\dagger$, with the result that
\begin{equation}
{\rm STr}'_{{\cal B}_E}\left(e^{-\Delta \tau}\right)={\rm STr}_{{\cal B}'_E}\left(e^{-\Delta \tau}\right).
\end{equation}
In general, however, it seems that we have to resort to the more complicated system
Eqs. (\ref{newg1}-\ref{newg3}).

We turn finally to the boundary conditions ${\cal B}_M$ which fix the field
$\gamma_{ij}$ on the boundary. A pure gauge mode would have to satisfy $s\gamma_{ij}=0$,
\begin{equation}
c_{(i|j)}+K_{ij}c=0.
\end{equation}
In the special case $K_{ij}=0$, we have solutions $c_i=0$ for any function $c$ on the boundary.
As before, these correspond to the bounded exponential solutions to the leading order system
of equations. However, for the more general case $K_{ij}=\kappa h_{ij}$, there are at most a finite
number of solutions corresponding to conformal killing vectors of ${\cal \partial M}$ and the origin
of the non-ellipticity in this case does not lie in the gauge modes. This case
has been investigated for spherical backgrounds \cite{Marachevsky:1995dr,Esposito:1995zf,Esposito:2012pi}, 
where it appears that the heat kernel diverges for points on the boundary, but nevertheless the generalised zeta 
function $\zeta(s)$ exists and can be analytically continued to $s=0$.
 
\section{Conclusion}

We have seen that non-ellipticity is a common feature in BRST-invariant boundary value problems 
and it can be associated in some cases with a residual gauge invariance. In these cases a well-defined 
set of boundary conditions can be obtained by adding extra tangential derivatives and this has
an interpretation in terms of extra gauge-fixing terms on the boundary. The reduced heat kernel, 
constructed by leaving out the gauge zero modes, is also consistent with BRST invariance and 
can be used to calculate one-loop phenomena in quantum gauge theories with boundaries. 
We found examples where this procedure gives results which are equivalent to mixed
BRST boundary value systems with no tangential derivatives.

A new set of BRST-invariant boundary conditions for quantum gravity which fix the
extrinsic curvature at the boundary has been found. These boundary conditions give rise to a 
well-defined heat kernel and can be used to construct the effective action. 
Boundary conditions on the extrinsic curbature are consistent with local supersymmetry and arise in 
supergravity theories \cite{Moss:2004ck,vanNieuwenhuizen:2005kg}. 

Although the addition of tangential derivatives formally restores strong-ellipticity, it does
necessarily allow for a non-problematic asymtotic expansion of the heat kernel
\cite{Barvinsky:2006pg}. This can affect quantum field theory divengences and
renormalisation of couplings. The heat-kernel asymptotics for the boundary conditions
discussed here is worth further investigation.

Applications of quantum gravity to quantum cosmology and the Hartle-Hawking state usually
have a fixed boundary geometry. The generalised zeta function exists when the background
is a sphere with boundary \cite{Marachevsky:1995dr,Esposito:1995zf,Esposito:2012pi}.  
However, there are at most a finite number of gauge 
zero modes in the BRST-invariant boundary value problem, and an understanding of 
non-ellipticity in this case remains elusive.

\appendix

\section{Tangential decompositions}

Consider a manifold ${\cal M}$ with boundary ${\cal \partial M}$, tangential vectors
$e_i{}^\mu$ and inward unit normal $n^\mu$. Indices $i$ and $n$ will denote projection 
in the tangential and normal directions
respectively. The covariant derivative `$;$' on ${\cal M}$ induces a covariant derivative `$|$' on
the boundary ${\cal \partial M}$ and the extrinsic curvature $K_{ij}=n_{i;j}$. The
normal derivive will be denoted by a dot, and we choose an extension of
the normal vector so that  $\dot n^\mu=0$ at the boundary.

\subsection{Covectors}
For a covector $A_\mu$, let
$A_i=e_i{}^\mu A_\mu$ and $A=n^\mu A_\mu$.
The components of the covariant derivatives are
\begin{eqnarray}
A_{i;j}&=&A_{i|j}+K_{ij}A,\\
A_{n;j}&=&A_{|i}-K_i{}^jA_j,\\
A_{i;n}&=&\dot A_i,\\
A_{n;n}&=&\dot A.
\end{eqnarray}
The gauge fixing function ${f}=A_\mu{}^{;\mu}$ is therefore
\begin{equation}
{f}=\dot A+A_i{}^{|i}+KA.
\end{equation}
For the Laplacian,
\begin{eqnarray}
A_{i;\mu}{}^{\mu}&=&\ddot A_i+A_{i|j}{}^j+K\dot A_i-KK_i{}^jA_j-K_i{}^j K_j{}^kA_k
+K_{ij}{}^{|j}A+K_i{}^jA_{|j}-\dot K_i{}^jA_j,\\
A_{n;\mu}{}^{\mu}&=&\ddot A+A_{|i}{}^i+K\dot A-K^{ij}A_{i|j}.
\end{eqnarray}

\subsection{Symmetric tensors}
For a symmetruc tensor $\gamma_{\mu\nu}$, let
$\gamma_{ij}=e_i{}^\mu e_j{}^\nu\gamma_{\mu\nu}$,  $\gamma_i=e_i{}^\mu n^\nu\gamma_{\mu\nu}$ and
$\gamma=n^\mu n^\nu\gamma_{\mu\nu}$.
The components of the covariant derivatives are
\begin{eqnarray}
\gamma_{ij;k}&=&\gamma_{ij|k}+2K_{k(i}\gamma_{j)},\\
\gamma_{in;j}&=&\gamma_{i|j}+K_{ij}\gamma-K_i{}^k\gamma_{kj},\\
\gamma_{nn;k}&=&\gamma_i-2K_i{}^j\gamma_j,\\
\gamma_{ij;n}&=&\dot\gamma_{ij},\\
\gamma_{in;n}&=&\dot\gamma_i,\\
\gamma_{nn;n}&=&\dot\gamma.
\end{eqnarray}
The decomposition of the gauge-fixing function ${f}_\mu=\overline\gamma_{\mu\rho}{}^{;\rho}$
into $f_i=e_i{}^\mu f_\mu$ and $f=n^\nu f_\mu$ is,
\begin{eqnarray}
{f}_i&=&\dot\gamma_i+K_i{}^j\gamma_j+K\gamma_i+\gamma_{ij}{}^{|j}
-\frac12\gamma^j{}_{j|i}-\frac12\gamma_{|i},\label{gfi}\\
{f}&=&\frac12\dot\gamma-\frac12\dot\gamma_i{}^i+\gamma_i{}^{|i}+K\gamma-K^{ij}\gamma_{ij},\label{gf}
\end{eqnarray}
The BRST transformations become
\begin{eqnarray}
s\gamma_{ij}&=&2(c_{(i|j)}+K_{ij}c),\label{sg1}\\
s\gamma_i&=&\dot c_i-K_i{}^jc_j+c_{|i},\\
s\gamma&=&2\dot c.\label{sg3}
\end{eqnarray}
When transforming a normal derivative, we use
\begin{equation}
(c_{i|j})\dot{\phantom a}=\dot c_{i|j}+(K_{ij}{}^{|k}-K^k{}_{(i|j)})c_k.
\end{equation}
For example,
\begin{equation}
s\dot\gamma_{ij}=2\dot c_{(i|j)}-2K_{(i}{}^k c_{j)|k}+2(K_{ij}{}^{|k}-K^k{}_{(i|j)})c_k
+2\dot K_{ij}c+2K_{ij}\dot c.
\end{equation}
The tangential decomposition relates to the standard canonical decomposition
of gravity in the following way. We write the metric in terms of an intrinsic
metric $h_{ij}$, lapse $N$ and shift $N^i$,
\begin{equation}
ds^2=N^2dt^2+h_{ij}(dx^i+N^i dt)(dx^j+N^jdt).
\end{equation}
In the fixed basis $e_i=\partial_i$ and $n^\mu=N^{-1}(\partial_t-N^i\partial_i)$, the metric fluctuations are
\begin{eqnarray}
\gamma_{ij}&=&(2\kappa)^{-1}\delta h_{ij},\\
\gamma_i&=&(2\kappa)^{-1}N^{-1}h_{ij}\delta N^j,\\
\gamma&=&(2\kappa)^{-1}2 N^{-1}\delta N.
\end{eqnarray}
The canonical momentum is given by
\begin{equation}
p^{ij}={\partial {\cal L}\over \partial \dot h_{ij}}=
{1\over 2\kappa}{\partial {\cal L}\over \partial \dot \gamma_{ij}}
\end{equation}
where ${\cal L}$ is the Lagrangian density with only first-order time derivatives
obtained from the Einstein-Hilbert action.

\bibliography{paper.bib}

\end{document}